\documentstyle[12pt]{article}
\begin{document}
\title{\LARGE \bf Sterile neutrinos and \\
\LARGE \bf
supernova nucleosynthesis}
\author{Juha T. Peltoniemi
\\
\normalsize \em Research Institute for Theoretical Physics, \\
\normalsize \em Box 9,
00014  University of Helsinki, Finland  \\
{\normalsize \sc Juha.Peltoniemi@Helsinki.FI}             }
\date{30 October 1995}
\maketitle
\begin{abstract}\normalsize
The role of the sterile neutrinos to the r-process nucleosynthesis
in supernova explosions is studied.
Previously it has been argued that a large part of neutrino mixing
can be excluded if the supernovae are the origin of the heavy elements.
It is shown that a conversion to sterile neutrinos may
evade those limits. The possibility that such conversions
can enhance the nucleosynthesis is investigated.
The desired mass spectrum is consistent with neutrino masses
suggested by other observed phenomena, like the solar neutrino problem,
the atmospheric neurino problem, dark matter and the LSND signals.
\end{abstract}
\vfill

\noindent
{\bf hep-ph/9511323   }\hfill HU-TFT-95-69\\

\noindent
{\sl A talk given at the Third Tallinn Symposium on Neutrino Physics,\\
7-11 October 1995, Lohusalu, Estonia}

\clearpage

\section{Light sterile neutrinos}

Many theories beyond the standard model involve
sterile neutrinos or similarly behaving objects.
Often the new neutral fermions are not called
neutrinos, since they may originate from completely different physics, but in
many occasions
they behave as sterile neutrinos, and may even mix with the known
neutrinos. Here   generically any  fermion
without standard model interactions, mixing with the ordinary neutrinos,
is called a sterile neutrino.

Typically the sterile neutrinos are very heavy, being hence out of
interest in astrophysics.
However, there are also several plausible ways to generate light sterile
neutrinos.
The small mass for the sterile neutrinos could be generated by
similar mechanisms as that of the ordinary neutrinos, like
see-saw or radiative mechanisms.
Since the basic theory does neither predict
nor forbid light sterile neutrinos,
the
existence of them is fundamentally an experimental problem.

Light sterile neutrinos can provide a solution to the solar neutrino
deficit, the atmospheric neutrino problem, the missing matter of
the universe, the anomalous ionization of interstellar hydrogen,
the explosion of a supernova,
the crisis of the big bang nucleosynthesis, and
the anomalies observed in the Karmen
experiment (See \cite{Peltoniemi95a} and references therein). Most of these
problems
can also be solved individually
without sterile neutrinos, either by ordinary neutrinos or by some other
objects. It is also possible that some, if not all, anomalies are due to
misconceptions in underlying theory
or uncontrolled phenomena in the experimental
apparatus.

The standard electroweak model with three massive
neutrinos allows only a simultaneous solution of two or three of the above
problems, unless excessive fine tuning is applied.
By introducing three sterile neutrinos one can solve almost all of the
above problems simultaneously, as well as fit the alleged neutrino mixing
observed by LSND \cite{LSND95}.

\section{Supernova nucleosynthesis}

It is widely believed thatmost
of the heavy elements observed in the universe,
as well as on the earth, have been produced in supernovae. The reactions
responsible for this are called the r-process.  After the collapse the
iron
mantle is at heavy pressure and high temperature, and
there are lots of free neutrons available. The free neutrons are captured by
nuclei which can grow to the edge of stability. Since the neutron capture is
faster than beta decay, the reactions can proceed through stability barriers.
The process lasts only a few tens of seconds, and the
products are ejected in the universe in the explosion of the star.

The success of the r-process relies on the neutron excess. If for some
reasons there were more protons than neutrons, no heavy nuclei
could be formed since there are no stable nuclear configuration with
the proton number exceeding the neutron number for such mass numbers.

Under the immense neutrino radiation the proton to neutron ratio is
dictated by the neutrino interactions. Even though the relevant region
is far outside the neutrinosphere, the neutrino radiation is so intensive
that all the nucleons are continuously bombarded by neutrinos. Hence,
to a  good approximation the neutron fraction is given by
\begin{equation}\label{p-n}
Y_{\mbox{n}} = \frac{1}{1+ \frac{F_{\nu_e}\langle E_{\nu_e}\rangle}
{F_{\bar{\nu}_e}\langle E_{\bar{\nu}_e}\rangle }
},
\end{equation}
where $F$ is the neutrino energy flux and $\langle E \rangle$ the average
neutrino energy. Assuming the fluxes to be sufficiently
equal, and the average energies to be
$\langle E_{\nu_e}\rangle \sim 11$ MeV,
and  $\langle E_{\bar{\nu}_e}\rangle \sim 16$ MeV,
one obtains $Y_{\mbox{n}} \simeq 0.6$
which satisfies qualitatively the requirement. More detailed numerical
calculations have shown that indeed the standard picture of the supernova
neutrinos gives a successful synthesis of the heavy nucleids. However,
there still is a small mismatch,
the models do not seem to yield enough of some
elements.

\section{Neutrino oscillations and the r-process}

Any change of the neutrino fluxes could upset the nucleosynthesis.
A conversion between neutrino flavors would change
the neutron fraction by changing
the average energies of neutrinos, even when
the neutrino fluxes
would remain equal.
Since muon and tau neutrinos (and antineutrinos) are emitted from a deeper
region of the star, due to their lack of charged current reactions,
they have
higher average energies,
above 20 MeV.
Hence an interchange $\nu_e \leftrightarrow \nu_\mu$
would lead to $E_{\nu_e} > E_{\bar{\nu}_e}$, and consequently
$Y_{\mbox{n}} < 0.5$. That would be intolerable.

To save the r-process one has to exclude a neutrino flavor conversion
between the emission of neutrinos and their passage through the iron rich
regions
\cite{Qianetal93}. This leads to a forbidden range in the
neutrino mixing plane which
covers the neutrino dark matter mass range, and
especially excludes the attractive scenario of having
a massive muon neutrino as hot dark matter and explaining simultaneously the
possible oscillation events in LSND.

It has been proposed that an inverted mass
spectrum ($m_{\nu_\mu} < m_{\nu_e}$) would solve the contradiction
\cite{FullerPrimackQian95}. 
In this case
there is no conversion between muon neutrinos and
electron neutrinos, instead muon antineutrinos would convert to electron
antineutrinos.  The resulting neutrino spectrum would be less favored
by the observations of SN1987A, although not necessarily ruled out.
Another unaesthetic feature of this scenario
is the fancy shape of the required
neutrino mass matrix.

\section{Conversions to sterile neutrinos}

The sterile neutrinos are not produced significantly in the core
\cite{KainulainenMaalampiPeltoniemi91}
unless they have relatively strong non-standard interactions.
Hence the neutrino flux
emitted from the neutrinosphere consists solely of the standard neutrinos.
Consequently, any conversion to sterile neutrinos only reduces the
effective neutrino flux.

As is clear from above, a detraction from the electron neutrino flux
enhances the r-process, while diminishing the antineutrino flux
endangers its success. The deficit of electron neutrinos may take
place directly, $\nu_e \to \nu_S$, or indirectly via other flavors:
$\nu_\mu \to \nu_S$ and then $\nu_e \to \nu_\mu$. The indirect
way is not necessarily more complicated.

A conversion from muon neutrinos to electron neutrinos can wash out
the effects of any previous transmutations between electron neutrinos
and sterile neutrinos. Hence the possibility for the transition
$\nu_e \to \nu_S$ does not evade the limits for the mixing between
active neutrinos, unless it occurs after the conversion $\nu_\mu \to \nu_e$.

The matter induced effective potentials for neutrinos are given by
\begin{eqnarray}\label{nuepot}
V(\nu_e) &=& V_0 (3 Y_e - 1 + 4Y_{\nu_e}\alpha ), \\
V(\nu_\mu) &=& V_0 (Y_e-1 + 2 Y_{\nu_e}\alpha)
\end{eqnarray}
where $Y$ are the relevant abundances of respective particles in
matter, and $V_0$ is a linear function of density.
The factor $\alpha$ ($0<\alpha<1$) appears in the free streaming
region since the neutrinos are all going
in the same direction out of the core,
and it is inversely
proportional to the square of the distance from the center.

In the region outside the core where $0.35 < Y_e < 0.5$, the relevant
potential differences for the different transitions are related as
\begin{equation}   \label{hie}
|\Delta V (\nu_e \to \nu_\mu)| > |\Delta V (\nu_e \to \nu_S)|
> | \Delta V (\nu_\mu \to \nu_S)|.
\end{equation}
However, the mass differences must also fit: for $\nu_e \to \nu_S$
the mass eigenstate dominated by the sterile state must be the
heaviest state, and for $\nu_e \to \nu_\mu$
and for $\nu_\mu \to \nu_S$ that by muon neutrinos.
For antineutrinos, i.e. right-handed Majorana states,
the mass differences must
be of the opposite sign.
If all transitions were described by sufficiently equal mass differences,
the one with the smallest potential
difference would cross the resonance first
(for an outgoing neutrino).

At regions close to the inner core the potentials may behave less
smoothly. In those regions the electron
 density drops due to the neutronisation
process, and can fall below 1/3. In such a case the potential for electron
neutrinos would change sign, and a resonance for the transition $\nu_e \to
\nu_S$ would occur for arbitrarily small
mass differences. Because of a specific evolution of the star, two level
crossings are formed, only one of
which may evolve outside the neutrinosphere
 \cite{Peltoniemi91a}.
The adiabaticity requirements are very stringent, and cannot be trivially
satisfied by light neutrinos ($m_\nu < 10$ eV) for any mixing. However, the
adiabaticity
condition, as well as the position of the level crossing may depend
strongly on the neutrino conversion itself, hence little exact can be said
at this stage. Even worse, for an inhomogenous or anisotropic case
there may be several resonances more.

To arrange a resonance for the transition $\nu_e \to \nu_S$ after the
respective conversion between electron and muon neutrinos a specific
mass hierarchy is required: $m_{\nu_e'} < m_{\nu_S'} < m_{\nu_\mu'}(1/2)$.
Again, to define the required range more exactly requires a thorough
numerical study which is out of the scope of this work.
Nevertheless, it can be said safely that the solar neutrino mass scale is
definitely too small for
this purpose.

In the presence of a conversion between electron and muon neutrinos
the transition from muon neutrinos to sterile neutrinos can as well
affect the nucleosynthesis \cite{Peltoniemi95a}. This possibility is
quite natural because of the above hierarchy (\ref{hie}).
Hence, whenever the mass differences are sufficiently equal,
the $\nu_\mu \to \nu_S$ transition
can naturally pass the resonance zone outside the neutrinosphere, but
before the resonance between electron and muon neutrinos.
This occurs for example in the scheme having two light neutrino
mass eigenstates made of the
electron neutrino
and a sterile neutrino, and a heavier neutrino
mass eigenstate made mainly of the
muon neutrino.
Hence, if the muon neutrino flux has been sufficiently reduced,
an interchange between muon and electron neutrinos would not
kill the r-process, instead it can even enhance it.

It can be estimated that
the transition $\nu_\mu \to \nu_S$ is adiabatic, for neutrino masses
in the dark matter range, if the mixing angle satisfies
\begin{equation}
\sin^2 2 \theta_\mu > 10^{-4} \ldots 10^{-3}.
\end{equation}
However, a partial transition is sufficient,
so that even lower mixings may be
enough.

The resonance zones ($\nu_\mu \to \nu_S$)
and ($\nu_\mu \to \nu_e$) are situated
very closely, and if they
are too wide they overlap. One can estimate that
the width of the resonances is narrower than
the distance
between the resonance zones
if
\begin{equation}
\sin^2 2\beta < 0.02,
\end{equation}
where $\beta$ is the largest mixing angle.
For the relevant mixing angles the overlap of the resonances does
not spoil the conversions.

\section{Discussion}

Although our understanding about the origin of the elements has made huge
progress in the recent years,
the birthplace of the heavy elements has not yet
been proven.  This work relies
on the assumption that the heavy element nucleosynthesis indeed
takes place in supernovae. This is a very justified assumption
since there are no other good candidates to accommodate that process.

It was found that light sterile neutrinos with a large mixing with the
ordinary neutrinos would affect
the r-process. Such mixings are not
forbidden by either laboratory experiments
or astrophysical arguments. A large mixing would, however, violate the
previous cosmological limits for the sterile neutrino mixing
\cite{EnqvistMaalampiSemikoz95}, but since
these limits are based on the currently unjustified \cite{Hataetal95}
requirement that
there are less than 0.4 new effective neutrinos, there is no reason
to abandon these solutions.

The introduction of sterile neutrinos provides a simple way to save the
interpretation of the identity of the hot dark matter and the LSND signal
as an indication of neutrino mass, without stretching too much the
theory. Hence, even though the interpretation of the Los Alamos neutrino
experiment
is at present premature, it has no imminent contradiction with astrophysics.

The required mass matrices leave the possibility to solve the solar and
atmospheric neutrino problems by neutrino oscillations. The simplest
solution compatible with everything would be the model with two very
light neutrinos, consisting of the electron neutrino and a sterile neutrino,
and two heavy neutrinos, made of muon and tau neutrinos in the few eV range.
The solar neutrino problem could
then be solved by the conversion from electron
neutrinos to sterile neutrinos, the atmospheric neutrino problem by the
oscillation between muon and tau neutrinos. These would also form the
hot dark matter. The mixing between electron and muon neutrinos would
then most naturally be in the region visible to LSND.

It is not yet clear whether the existence
of sterile neutrinos could actually
improve the fit of the nucleosynthesis model
to the observations. Qualitatively
this could happen, since a conversion to
sterile neutrinos could increase the
neutron abundance. However, the full
nucleosynthesis is a very complicated
network, and modelling it requires a detailed
numerical simulation which has not
been done yet for this case.

\setlength{\itemsep}{0mm}


\begin{thebibliography}{10}

\bibitem{Peltoniemi95a}
J.~T. Peltoniemi, hep-ph/9506228 (unpublished).

\bibitem{LSND95}
C. Athanassopoulos {\it et~al.}, Phys.~Rev.~Lett. {\bf 75}, 2650  (1995).

\bibitem{Qianetal93}
Y.-Z. Qian {\it et~al.}, Phys.\ Rev.\ Lett. {\bf 71},  1965  (1993);
Y.-Z. Qian and G.~M. Fuller, Phys.\ Rev.\ D {\bf 51},  1479  (1995).

\bibitem{FullerPrimackQian95}
G.~M. Fuller, J. Primack, and Y.-Z. Qian, Phys.\ Rev.\ D {\bf 52},  1288
  (1995);
G. Raffelt and J. Silk,  hep-ph/9502306 (unpublished);
D. Caldwell and R.~N. Mohapatra, Phys.\ Lett.\ B {\bf 354},  371  (1995).

\bibitem{KainulainenMaalampiPeltoniemi91}
K. Kainulainen, J. Maalampi, and J.~T. Peltoniemi, Nucl.\ Phys. {\bf B358},
  435  (1991).

\bibitem{Peltoniemi91a}
J.~T. Peltoniemi, Astron. Astroph. {\bf 254},  121  (1992).

\bibitem{EnqvistMaalampiSemikoz95}
K. Enqvist, J. Maalampi, and V.~B. Semikoz,  hep-ph/9505210
  (unpublished).

\bibitem{Hataetal95}
N. Hata {\it et~al.},  hep-ph/9505319 (unpublished).

\end{thebibliography}
\end{document}